\documentclass{PoS}

\usepackage{graphicx}
\usepackage{amsmath}
\usepackage{amsfonts}
\usepackage{amssymb}
\usepackage{mathrsfs}
\usepackage{bm}
\usepackage{verbatim}

\newcommand{\beq}{\begin{equation}}
\newcommand{\eeq}{\end{equation}}
\newcommand{\bea}{\begin{eqnarray}}
\newcommand{\eea}{\end{eqnarray}}

\newcommand{\reffig}[1]{FIG.~\ref{#1}}

\newcommand{\tr}[1]{\text{#1}}

\newcommand{\iv}[1]{\bm{#1}}

\newcommand{\mqcd}{M_{\textrm{QCD}}}

\title{Pion-nucleon scattering around the delta resonance}

\ShortTitle{Pion-nucleon scattering around the delta resonance}

\author{\speaker{Bingwei Long}\\
        European Centre for Theoretical Studies in Nuclear Physics and
        Related Areas (ECT*)\\
        Strada delle Tabarelle 286\\
        I-38100 Villazzano (TN), Italy\\
        E-mail: \email{long@ect.it}}

\abstract{
We develop a generalized version of
heavy-baryon chiral perturbation theory to describe 
pion-nucleon scattering in a kinematic
domain that 
extends continuously from threshold to the delta-isobar peak. 
The $P$-wave phase shifts are used to illustrate this framework.
}

\FullConference{6th International Workshop on Chiral Dynamics\\
                 July 6-10 2009\\
                 Bern, Switzerland}

\begin{document}

\section{Introduction}
A prominent feature of pion-nucleon ($\pi N$) scattering is 
the delta resonance, $\Delta(1232)$, a peak in the elastic
cross section at the center-of-mass (CM) energy 
$m_{\Delta} \equiv m_N + \delta \simeq 1230$ MeV,
where $\delta \sim 290$ MeV is the nucleon-delta mass splitting
\cite{pdg}. A resonance can be studied by
considering the unitarity and analyticity of the $S$ matrix; however,
the accuracy is hard to improve systematically with these general
principles alone. Our goal here is to
investigate $\pi N$ scattering from threshold up to the delta resonance
in an effective field theory (EFT) (for more details, see
Ref.~\cite{bwlbira-delta}). 

Following several seminal papers \cite{weinberg79},
EFTs have been developed as model-independent
approximations to low-energy strong interactions, which can be
systematically improved by a series in powers of $Q/\mqcd$, where $Q$
refers generically to small external momenta and $\mqcd \sim 1$ GeV
is the characteristic QCD scale. For reviews, see, for example,
Refs.~\cite{BerKaiMei,bira99review}. Chiral perturbation theory (ChPT)
specializes in processes involving at most one nucleon \cite{BerKaiMei}.
ChPT with only pion and nucleon fields 
has been extensively applied to near-threshold $\pi N$ scattering
\cite{threshold},
resulting in a perturbative expansion in powers of $Q/\delta$ and 
$m_\pi/\delta$, which converges slowly as $m_\pi/\delta \simeq 1/2$. The
convergence can be
improved with an explicit delta field. The explicit delta in $\pi N$
scattering within standard ChPT has been explored
\cite{threshold-delta1} and demonstrated in a fully consistent
calculation \cite{threshold-delta2}.

Nevertheless, the perturbative nature of standard ChPT makes it impossible to
describe the delta resonance, a non-perturbative phenomenon. A
non-perturbative treatment of the delta within ChPT was considered in
Ref.~\cite{ellis-tang}; however, a systematic resummation did not
exist until the seminal work of Ref.~\cite{pascalutsa}, where it was
justified by a power counting 
based on three separate scales $m_\pi \ll \delta \ll \mqcd$, and this
idea has been applied to various electromagnetic
reactions in the delta region,
but for $\pi N$ scattering few results have been published 
\cite{morepascalutsa2}.

We employ a power counting developed for generic 
narrow resonances \cite{resonances}, in which there are only
two scales $M_{\tr{lo}} \sim \delta \sim m_\pi$
and $M_{\tr{hi}} \sim \mqcd$. Thus the EFT expansion of the $\pi N$
scattering amplitude pursued here is in powers
of $Q/M_{\tr{hi}}$ and $M_{\tr{lo}}/M_{\tr{hi}}$. 
The kinematic region under consideration spans over both threshold and the
resonance.

\section{Effective Lagrangian}

To establish the notation, we review how the effective 
Lagrangian is constructed (for more details of building chiral
Lagrangian, see \textit{e.g.}, Refs.~\cite{BerKaiMei,
  bira99review, weinbergbook, bira-thesis}). The effective Lagrangian
should inherit the symmetries of QCD: Lorentz invariance,
(approximate) two-flavor chiral symmetry ($SU(2)_L \times
SU(2)_R$), parity, time-reversal invariance, and baryon-number
conservation.

In the kinematic region where the EFT holds, external
momenta are much smaller than the nucleon mass, $Q \ll m_N$, and thus
Lorentz invariance can be fulfilled perturbatively in powers of
$Q/m_N$. One can start with a relativistic Lagrangian using the
Rarita-Schwinger field for the delta, and then reduce from it its nonrelativistic
version \cite{pascalutsa, hemmertdelta}. This way, however, extra effort needs to
be taken in order to control the spurious spin-1/2
sectors of the Rarita-Schwinger field. We employ another approach that
starts with heavy-baryon fields $N$ for
the nucleon and $\Delta$ for the delta, which are, respectively,
two- and four-component spinors in spin and
isospin spaces. Eventually, the effective Lagrangian only has
the baryon degrees of freedom that represent forward propagation. The
crucial ingredient in this approach is to develop an order-by-order Lorentz
transformation, by which one can constrain the coefficients of the
rotation-invariant operators \cite{bira-thesis, bw-nonrel}.

Due to the presence of the delta field $\Delta$, one needs $2 \times 4$
matrices $\vec{S}$ in spin space to make a three-vector $N
\Delta$ bilinear, and $\Omega_{ij}$ a three-tensor. Similar transition
matrices, $\iv{T}$ and $\Xi_{a b}$, can be 
defined in isospace.

The chiral-invariant operators are isoscalars that are made of pion
covariant-derivative $\iv{D}_\mu \equiv D^{-1} \partial_{\mu}
\iv{\pi}/2 f_\pi$ with $D \equiv 1 + \iv{\pi}^2/4f_\pi^2$, $N$,
$\Delta$, and their covariant derivatives, for 
example, $\mathscr{D}_{\mu}\Delta \equiv \left(\partial_{\mu} +
  \iv{t}^{\left(\frac{3}{2}\right)}
\bm{\cdot} \iv{E}_{\mu} \right) \Delta$ with $\iv{E}_{\mu} \equiv i\,
\iv{\pi}/f_\pi \bm{\times} \iv{D}_{\mu}$.

We use the so-called chiral index $\nu$ \cite{weinberg79} to organize
the operators of the effective Lagrangian $\nu = d + m + n_{\delta} +
f/2 - 2$, where $d$, $m$, $n_\delta$, and $f$ are the numbers of
derivatives, powers of $m_{\pi}$, powers of $\delta$, and fermion
fields, respectively. In constructing the Lagrangian, we use 
integration by parts and field redefinitions to remove time derivatives on
baryon fields except for the kinetic terms. The 
Lagrangian terms
with the two lowest indices are given by \cite{bira-thesis}
\bea
\mathcal{L}^{(0)} & = & 2 f_\pi^2 \iv{D}^2
- \frac{1}{2D} m_{\pi}^2 \iv{\pi}^2 
+  N^{\dagger} i \mathscr{D}_0 N 
+ g_A N^{\dagger} \iv{\tau} \vec{\sigma} N \bm{\cdot} \cdot \vec{\iv{D}} 
\nonumber \\ 
& & + \Delta^{\dagger} \left( i \mathscr{D}_0 - \delta \right) \Delta
+ 4 g^{\Delta}_A \Delta^{\dagger} \iv{t}^{(\frac{3}{2})}
\vec{S}^{(\frac{3}{2})} \Delta \bm{\cdot} \cdot \vec{\iv{D}} 
+ h_A \left( N^{\dagger} \iv{T} \vec{S} \Delta+ H.c. \right) 
      \bm{\cdot} \cdot \vec{\iv{D}} 
+\cdots  \label{eqn:lag0} 
\eea
and
\beq
\mathcal{L}^{(1)} = \frac{1}{2 m_N}  
\left( N^{\dagger} \vec{\mathscr{D}}^2 N
+ \Delta^{\dagger} \vec{\mathscr{D}}^2 \Delta \right)
- \frac{h_A}{m_N} 
 \left( i N^{\dagger} \iv{T} \vec{S}\cdot\vec{\mathscr{D}}\Delta + H.c.\right) 
 \bm{\cdot}\iv{D}_0 + \cdots \; , \label{eqn:lag1}
\eeq
while the next-higher index yields
\bea
\mathcal{L}^{(2)} &=& 
-\frac{\delta}{2 m_N^2} \Delta^{\dagger} \vec{\mathscr{D}}^2 \Delta 
+ \frac{h_A}{2 m_N^2} 
\left[
     \left(N^{\dagger} \iv{T} \vec{S}\vec{\mathscr{D}}^2 \Delta
   - 
N^{\dagger} \iv{T}\vec{S}\cdot\vec{\mathscr{D}}\vec{\mathscr{D}}\Delta\right)
     + H.c.   
\right] \bm{\cdot} \cdot\vec{\iv{D}} 
\nonumber \\
& &
+ \frac{h_A}{8 m_N^2} 
\left[
  \left(\delta_{lm} N^{\dagger} \iv{T}\vec{S}\cdot\vec{\mathscr{D}}\Delta 
 +3  N^{\dagger} \iv{T}S_l\mathscr{D}_m \Delta 
 +2 \epsilon_{ijl}  
    N^{\dagger}\iv{T}\Omega_{im} \mathscr{D}_j\Delta  \right) + H.c.
\right]
  \bm{\cdot} \mathscr{D}_l\iv{D}_m 
\nonumber \\
& &+ d_1 \left(N^{\dagger} \iv{T} \vec{S} \Delta+ H.c. \right)
      \bm{\cdot}\cdot \vec{\mathscr{D}}^2 \vec{\iv{D}}
+ d_2\, \frac{m_\pi^2}{D} 
   \left(1 - \frac{\iv{\pi}^2}{4f_\pi^2}\right) 
\left( N^{\dagger}\iv{T}\vec{S}\Delta + H.c.\right)\bm{\cdot}\cdot\vec{\iv{D}}
+ \cdots 
\label{eqn:lag2}
\eea
Here, $g_A$ ($g^\Delta_A$) is the $\nu=0$ axial-vector coupling of the nucleon
(delta)
and $h_A$ ($d_{1,2}$) is (are) the $\nu=0$ ($\nu =2$) $\pi N \Delta$ 
coupling(s). 

\section{Power counting}

When the CM (heavy-baryon) energies $E$ are much below the delta peak,
the power counting
is standard \cite{weinberg79, BerKaiMei, bira99review} 
with the simple generalization that $\delta$ counts as $Q$.
The contribution of a diagram with $A$ nucleons (here $A = 1$),
$L$ loops, and $V_i$ vertices
with chiral index $\nu_i$ is proportional to $Q^\rho$, with
\beq
\rho = 2 - A + 2L + \sum_i V_i \nu_i \; .
\label{eqn:std-con}
\eeq

\begin{figure}
\centering
\includegraphics[scale=1.1]{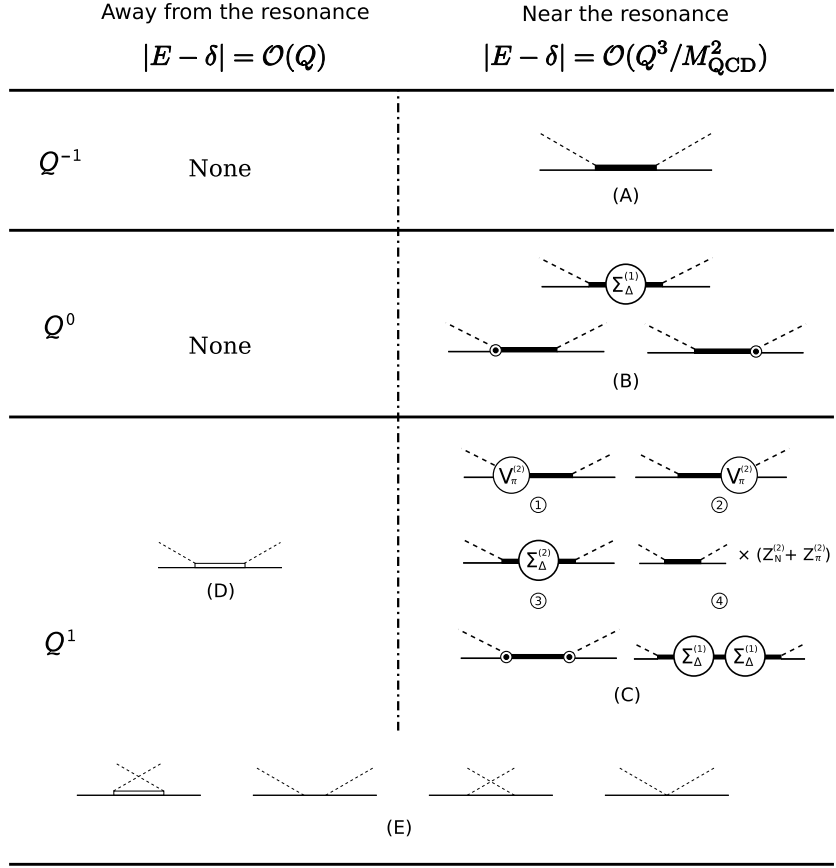}
\caption{\label{fig:powct}Contributions to $\pi N$ scattering 
up to order $Q^1$: 
(A) $Q^{-1}$ pole diagram; 
(B) $Q^{0}$ pole diagrams;
(C) $Q^{1}$ pole diagrams;
(D)\&(E) $Q^1$ tree diagrams,
of which (E) apply to both regions. $\Sigma_\Delta^{(n)}$ is the
$n$-th order delta self-energy, $V_\pi^{(n)}$ the $\pi N \Delta$
vertex function, and $Z_N^{(n)}$ ($Z_\Delta^{(n)}$) the nucleon
(delta) field renormalization constant.
}
\end{figure}

However, in the small region spanning the delta peak whose size is of the
leading-order (LO) delta self-energy, $|E - \delta| \sim
\Sigma_\Delta^{(0)}=\mathcal{O}(Q^3/\mqcd^2)$, a resummation is needed
in one-$\Delta$-reducible diagrams because one insertion of
$\Sigma_\Delta^{(0)}$ and the bare delta propagator
contributes $\mathcal{O}(1)$: $\Sigma_\Delta^{(0)}/(E - \delta)
 = \mathcal{O}(1)$. The resummation thus amounts to
a dressed propagator
\bea
S_\Delta^{(0)}(E)=
\left[E-\delta+\Sigma_\Delta^{(0)}(\delta)\right]^{-1} \; ,
\label{eqn:dressedp}
\eea
which scales as $\mqcd^2/Q^3$.
This is an enhancement of two powers over the generic situation.
As a consequence, 
in one-$\Delta$-irreducible diagrams the standard ChPT power counting 
\eqref{eqn:std-con}
still applies;
dressed delta propagators only need to be included in 
one-$\Delta$-reducible diagrams.
We thus arrive at a new power counting for one-$\Delta$-reducible
diagrams within a narrow window around the delta peak,
\beq
\rho = 2 - A - 2 n_\Delta + 2L + \sum_i V_i \nu_i \; ,
\label{eqn:rho2}
\eeq
where $n_\Delta$ is the number of dressed delta propagators. 
This is the non-electromagnetic version of $\rho$ derived
in a slightly different power counting in Ref.~\cite{pascalutsa}.
Diagrams up to $Q^1$ are listed in \reffig{fig:powct}.

It seems that the two different power-counting schemes, which are
applicable in two different regions, would lead to an EFT amplitude in
the form of a piecewise function in the energy. Even worse, separating
these two regions is somewhat arbitrary. A piecewise EFT is actually
unnecessary if we enforce the pole diagrams even in the off-the-pole
region, which is equivalent to shifting a subset of higher diagrams
into lower orders, \textit{i.e.}, a rearrangement of diagrams. This
sort of rearrangement
still retains the essence of the original power counting as long as 
one does not claim a higher accuracy by doing so.

\section{$\pi N$-scattering $T$ matrix}

The partial-wave $T$ matrix is related to the phase shifts, in the
channel with total
angular momentum $j$, orbital angular momentum $l$, and isospin $t$, by
\beq
T_{jlt}(E) \equiv -i \left\{\exp\left[ 2i\theta_{jlt}(E) \right] - 1\right\} 
\; .
\label{eqn:ps-def}
\eeq
In the following we will use
a more conventional notation for a specific partial wave: $l_{2t, 2j}$. For
example, $P_{13}$ refers to the $l=1$ ($P$ wave), $t=1/2$, and
$j=3/2$.

Here the exact relation between 
$E$ and the CM momentum $k$, $E = (m_N^2 + k^2)^{1/2} + (m_\pi^2 +
m_N^2)^{1/2} - m_N$,  is assumed,
meaning that certain trivial, kinematic $k/m_N$ 
terms are resummed ---what we refer to as semi-resummation. A strict
heavy-baryon expansion can be readily obtained afterwards.

At LO ($Q^{-1}$) there is only a pole diagram,
\reffig{fig:powct}(A), which contributes only to the $P_{33}$ wave,
\beq
T_{P_{33}}^{\text{LO}} = -\frac{\gamma^{(0)}(\delta)}{E - \delta +
  i\gamma^{(0)}(\delta)/2} 
\left[ 1 +\mathcal O\left(\frac{Q}{\mqcd}\right) \right] \; ,
\label{eqn:TP33LO}
\eeq
where
\beq
\gamma^{(0)}(\delta) =
\frac{h_A^2}{24 \pi f_\pi^2} \left(\delta^2 -m_\pi^2\right)^{\frac{3}{2}}
\left[1+ \delta/m_N + (\delta^2 -m_\pi^2)/(2m_N)^2\right]^\frac{3}{2} 
\frac{1+ \delta/m_N + (\delta^2 -m_\pi^2)/2m_N^2}{\left(1+\delta/m_N\right)^5}
\; .
\label{eqn:gamma0}
\eeq

The NLO ($Q^0$) amplitude has the same form as LO,
\beq
T_{P_{33}}^{\text{NLO}} = 
-\frac{\gamma^{(0)}(\delta)+\gamma^{(1)}(\delta)}{E - \delta +
  i\left[\gamma^{(0)}(\delta)+\gamma^{(1)}(\delta)\right]/2} 
\left[ 1 +\mathcal O\left(\frac{Q^2}{\mqcd^2}\right) \right] \; .
\label{eqn:TP33NLO}
\eeq
However, $\gamma^{(1)}(\delta)$ vanishes in the CM frame when we do not expand
kinematic relations in powers of $\delta/m_N$.

Summing up the pole (\reffig{fig:powct}(C)) and tree
(\reffig{fig:powct}(E)) diagrams, one first finds the NNLO amplitude
in the $P_{33}$ channel,
\beq
T_{P_{33}}^{\text{NNLO}} = 
 -\frac{\Gamma(E)}{E - \delta + i\Gamma(E)/2} \left[1 +i T_B(E)\right] 
+ T_B(E) + \mathcal
O\left(T_{P_{33}}^\text{LO}\,\frac{Q^3}{\mqcd^3}\right) \; ,
\eeq
where
\beq
T_B(E) = \frac{k^3}{6\pi f_\pi^2}\left(\frac{g_A^2}{E} +
  \frac{1}{36}\frac{h_A^2}{E+\delta} \right)
\quad \text{and} \quad
\Gamma(E) =\frac{\left(m_N^2 + k^2\right)^{1/2}}{E + m_N}\,
\frac{\left[h_A(1+\varkappa)\right]^2}{24 \pi f_\pi^2}\,k^3
\label{eq:B(E)}
\eeq
with
\beq
\varkappa \equiv \frac{k_\delta^2}{(4\pi f_\pi)^2} 
\left[\frac{(4\pi f_\pi)^2}{h_{A}} 
\left(-d_{1}+d_{2}\frac{m_\pi^2}{k_\delta^2} 
\right)
+ \mathrm{Re} \mathcal{G}(m_\pi/k_\delta)
\right]\; , 
\label{eqn:ftilde}
\eeq
where
\bea
\mathcal{G}(x) &=& \frac{2}{3}\left(1+x^2\right)^{-\frac{1}{2}} 
\left\{
-\pi\left(g_A^2 - 
\frac{81}{16} {g_A^\Delta}^2\right)
x^3
+ 2\pi i \left(g_A^2 + \frac{1}{72}h_{A}^2\right) 
\right. \nonumber \\
&& \left.
+  
\left[g_A^2 -\frac{1}{72}h_{A}^2 \left(13+15 x^2\right)+
\frac{81}{16} {g^\Delta_A}^2 \right]
\ln\left(\frac{\sqrt{1+x^2}-1}{\sqrt{1+x^2}+1}\right) 
\right\} \; 
\eea
and $k_\delta$ satisfies $\delta = (m_N^2 + k_\delta^2)^{1/2} + (m_\pi^2
+ k_\delta^2)^{1/2} - m_N$. 
Other channels are easy to calculate from the
one-$\Delta$-irreducible tree diagrams in \reffig{fig:powct}(E).
For the remaining $P$-wave channels,
\beq
T_{P_{13}}^{\text{NNLO}} = T_{P_{31}}^\text{NNLO} 
=\frac{1}{4} T_{P_{11}}^\text{NNLO}=
-\frac{k^3}{12 \pi f_\pi^2}
\left(\frac{g_A^2}{E} - \frac{2}{9}\frac{h_A^2}{E+\delta}\right)
\left[1 + \mathcal O\left(\frac{Q}{\mqcd}\right) \right]
\; . \label{eqn:Tp131} 
\eeq

\section{$P$-wave phase shifts}

\begin{figure}
\centering
\includegraphics[scale=0.75]{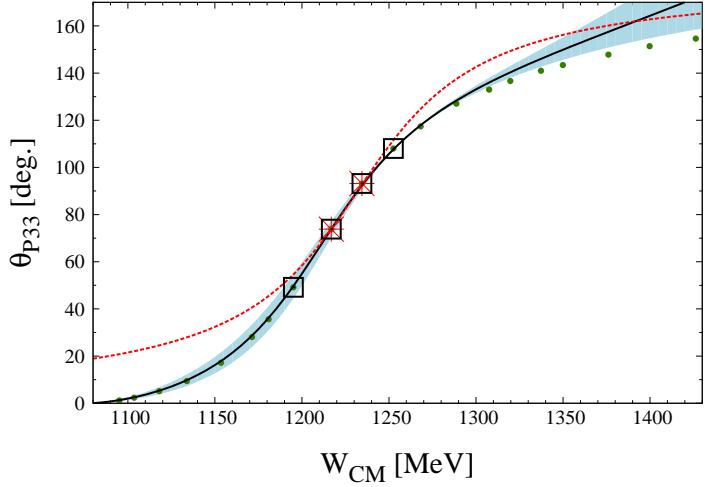}
\caption{\label{fig:p33}$P_{33}$ phase shifts (in degrees) as a function
  of $W_{\text{CM}}$ (in MeV), the CM energy including the nucleon mass.  
The EFT strict heavy-baryon expansion
at LO (NNLO)
is represented by the red dashed (black solid) line. 
The NLO curve coincides with LO.
The light-blue
band outlines the estimated systematic error of the NNLO curve.
The green dots are the
results of the GW phase-shift analysis \cite{gwpwa}. 
Points marked by a red star (black square) are inputs for LO (NNLO).
}
\end{figure}

\begin{figure}
\centering
\includegraphics[scale=0.55]{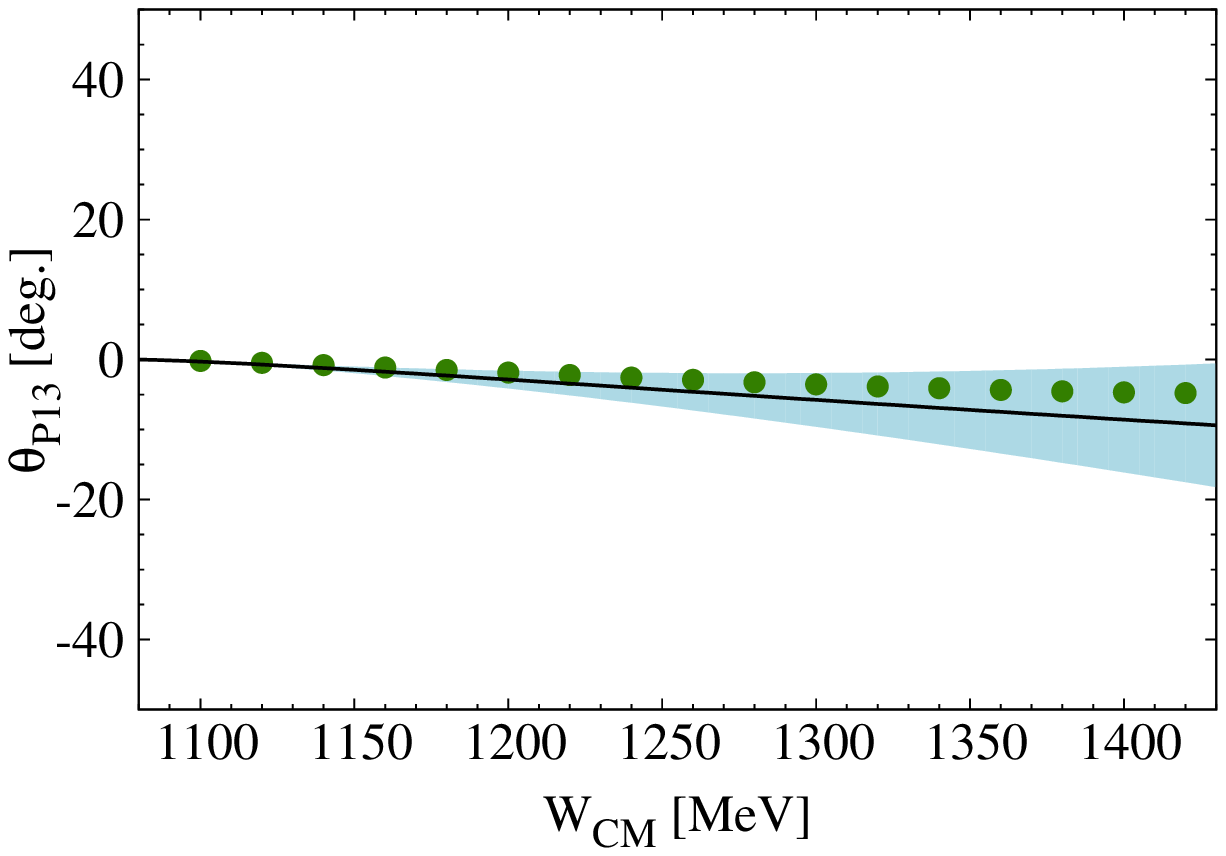}
\includegraphics[scale=0.55]{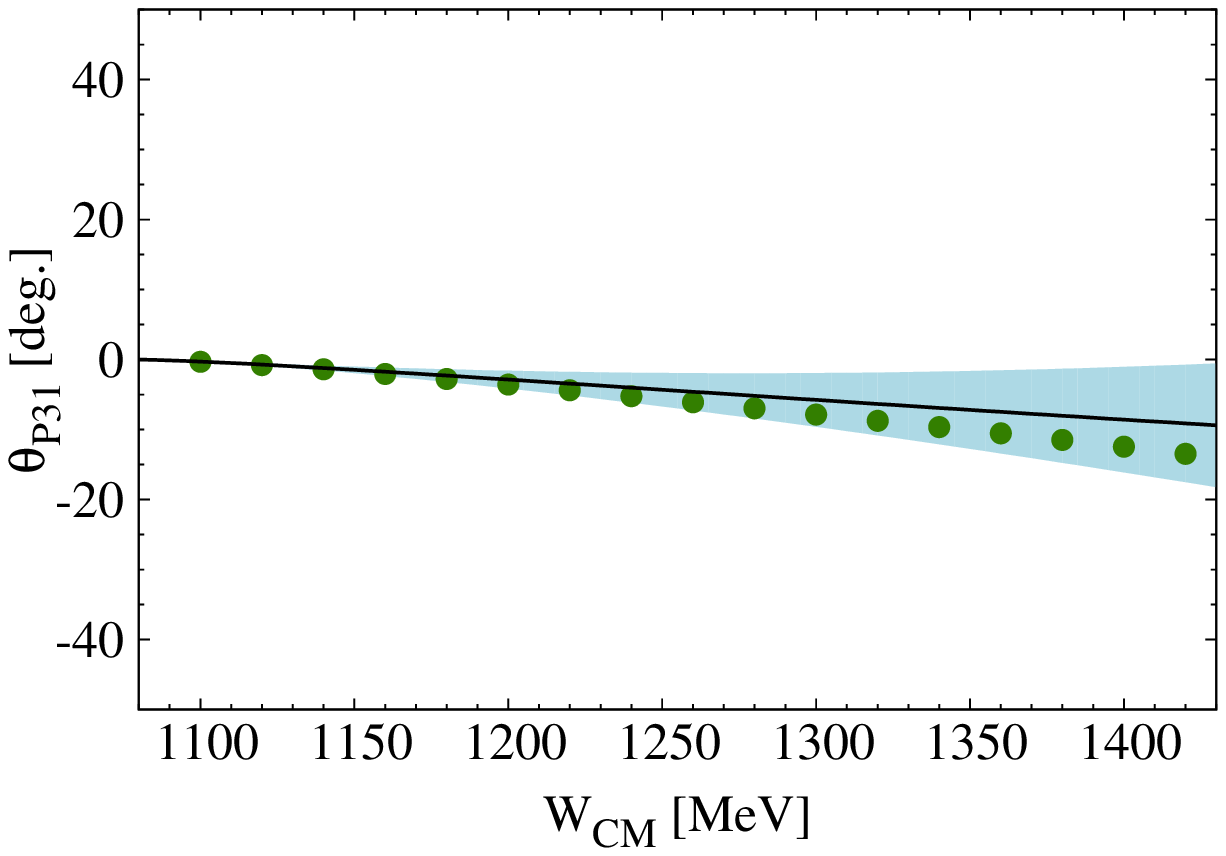}
\includegraphics[scale=0.55]{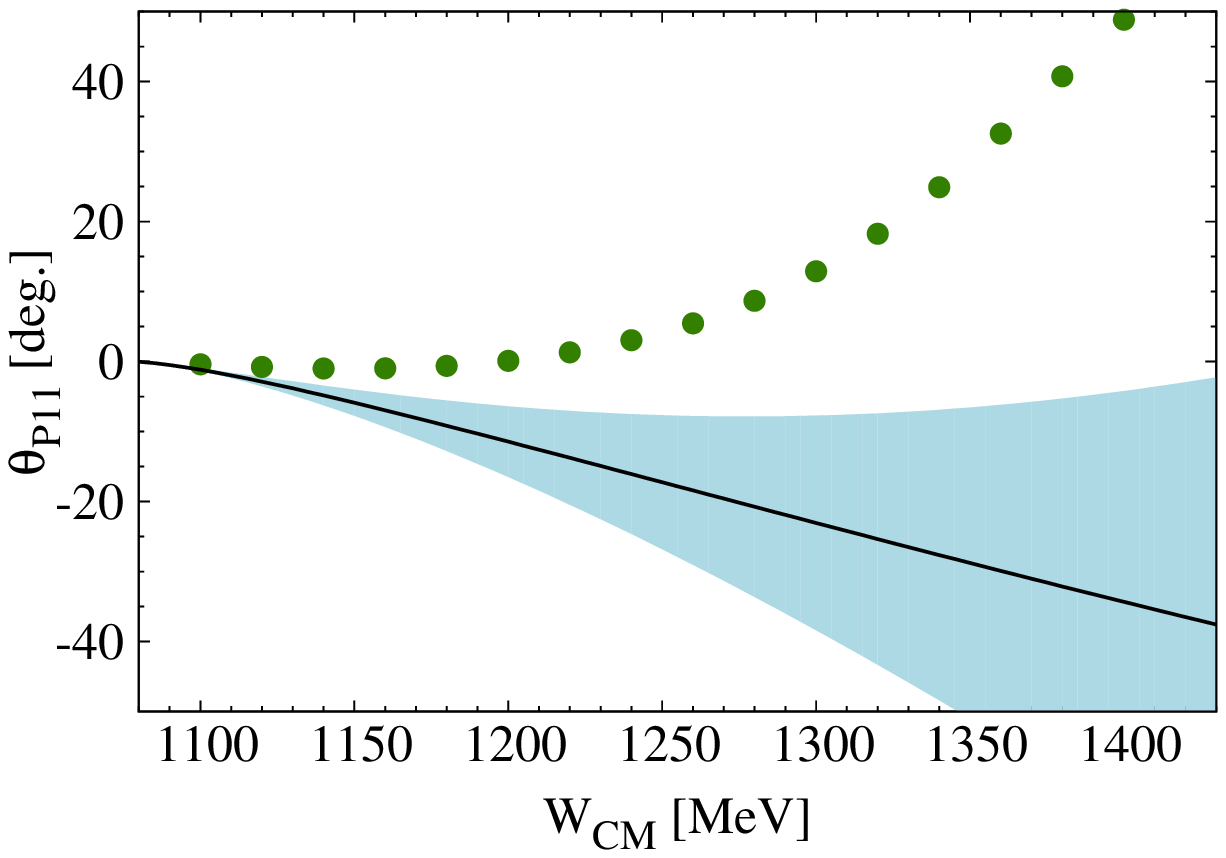}
\caption{\label{fig:otherpwaves} Predicted phase shifts  (in degrees)
in the $P_{13}$, $P_{31}$, and $P_{11}$ channels
as functions of $W_{\text{CM}}$ (in MeV),
the CM energy including the nucleon mass. 
LO and NLO vanish in these channels;
NNLO EFT results in the strict heavy-baryon expansion
are given by the black solid lines. 
The light-blue bands outline the estimated systematic errors of 
the NNLO curves.
The green dots are the
results of the GW phase-shift analysis \cite{gwpwa}. 
}
\end{figure}

A number of low-energy constants (LECs) can be determined from other processes,
such as pion decay and neutron decay. We adopt the following values: 
$m_\pi=139$ MeV, $m_N=939$ MeV, 
$g_A = 1.26$, and $f_\pi=92.4$ MeV. Our strategy of fitting is to
determine the free parameters, $\delta$,
$h_A$, and $\varkappa$ from the $P_{33}$ phase shifts around the delta
peak and then predict the phase shifts at lower energies in all
$P$ waves. Shown in FIGs.~\ref{fig:p33} and \ref{fig:otherpwaves} are
the EFT curves (strict heavy-baryon expansion used) fitted to the
partial-wave analysis (PSA) by the 
George Washington (GW) group \cite{gwpwa}. The PSA points used to
determine the free parameters are explicitly marked. Based on the
power counting we use, systematic errors of the EFT curves can be
estimated, shown in FIGs.~\ref{fig:p33} and \ref{fig:otherpwaves} as
light-colored bands.

The LECs extracted from the $P_{33}$ fit are given in 
TABLE \ref{tbl:lecmix}. One can estimate the errors in the NNLO values 
as the variation in each LEC within which the
NNLO $P_{33}$ curve in \reffig{fig:p33} roughly stays within the
error band. This way we find $\delta$/MeV, $h_A$, and $\varkappa$ to be within
$\sim \pm 4$, $\pm 0.30$, and $\pm 0.030$, respectively, of 
the NNLO values in TABLE \ref{tbl:lecmix}.

\begin{table}
\caption{\label{tbl:lecmix}Low-energy constants
extracted at LO, NLO, and NNLO from the fits using 
the strict heavy-baryon expansion.
}
\centering
\begin{tabular}{|c c c|c c c| c|}
\hline
\multicolumn{3}{|c|}{$\delta$ (MeV)}& \multicolumn{3}{c|}{$h_A$} &
$\varkappa$\\
 LO &NLO & NNLO & LO &NLO & NNLO & NNLO \\
\hline
\hline
293 & 293 & 321 & 1.98 & 4.21 & 2.85 & 0.046 \\
\hline
\end{tabular}
\end{table}

\section{Summary}

We have extended standard ChPT to deal with the non-perturbative delta
resonance in an EFT framework. The delta is treated as a
nonrelativistic particle from the beginning, rather than being
represented by the Rarita-Schwinger field.

Like other EFTs that deal
with non-perturbative phenomena, 
ours captures the non-perturbative structure in LO.
Subsequently, the power counting leads to a systematic,
perturbative improvement beyond LO.
We applied this power counting to low-energy $\pi N$
scattering, where we built the 
amplitudes up to NNLO. We fitted our $P$-wave 
amplitudes to the phase shifts given by Ref.~\cite{gwpwa}. 
With just three free
parameters, we obtained a good fit in the $P_{33}$ channel.

The EFT approach presented here also provides the basis for a model-independent,
unified description, from threshold to past the delta
resonance without discontinuity, of reactions involving other probes and targets,
including nuclei.

\end{document}